\documentclass[twoside,11pt]{article}
\usepackage{jmlr2e}

\jmlrheading{x}{2011}{x-xx}{5/11}{xx/11}{Grigory Pivovarov and Sergei Trunov}

\ShortHeadings{Text Clustering and Classification with Modularity}{Pivovarov and Trunov}
\firstpageno{1}

\begin{document}

\title{Clustering and Classification in Text Collections Using Graph Modularity}

\author{\name Grigory Pivovarov \email gbpivo@ms2.inr.ac.ru \\
       \addr Institute for Nuclear Research\\
       Russian Academy of Sciences\\
       Moscow, 117312, Russia
       \AND
       \name Sergei Trunov \email trunov7@gmail.com \\
       \addr Institute for Institutional Analyses\\
       Higher School of Economics\\
       Moscow, 109028, Russia}

\editor{Not Known}

\maketitle

\begin{abstract}
A new fast algorithm for clustering and classification of large collections of text documents is introduced. 
The new algorithm employs the bipartite graph that realizes the word-document matrix of the collection. Namely, the modularity of the bipartite graph is used as the optimization functional. Experiments performed with the new algorithm on a number of text
collections had shown a competitive quality of the clustering (classification), and a record-breaking speed.
\end{abstract}


\section{Introduction}

We explore a possibility of clustering (or classification) of documents. Clustering and classification are methods for information retrieval (for a recent review see \citet{Berry:2003}). The possibility we explore consists in combining two ideas considered previously. 

The first idea is co-clustering \citep{Dhillon:2001}, \citep{Zha:2001}. Co-clustering clusters along with the documents the words used in the documents. As an outcome, clusters of documents are generated along with corresponding clusters of words. This approach features the following advantages: Clusters of words generated as a byproduct of the approach can be used for interpretation of the clusters of documents; In the classification tasks, it is possible to use in the training sets separate words along with documents. The standard algorithm used within the co-clustering approach to reach the result is the spectral clustering \citep{Luxburg07}. (Computationally, spectral clustering finds eigenvectors of the graph laplacian. With a number of tricks, the eigenvectors are used for clustering.)

The second idea is modularity \citep{Newman-2006}. Modularity is a class of optimization functionals introduced in the studies of graph clustering. Let us compare modularity to other optimization functionals appearing within the widely used approach to clustering based on generative models \citep{ZhongG05}. These optimization functionals are various ``distances'' between the data and the model. Optimization consists in finding parameters of the model yielding the minimal distance. In contrast, the modularity is optimal when a ``distance'' between the data and a null-model is maximal. The null-model is a key notion for the modularity idea. The null-model models the data without structure (the most random data). Concretely, modularity is defined as follows. A functional on graph partitionings is picked out. Modularity is an additive or multiplicative difference of the value the functional takes on the graph under study and the mean value it takes on the null-model. From the above comparison we conclude that using modularity is in a way less demanding than using generative models, because it is easier to model randomness than specific data. Comparison of modularity-based approaches and generative models approaches is attempted in \citep{Newman-2011}, \citep{Bickel09}.     

Modularity has been used in text clustering \citep{Grinev:2009}. In this attempt, a dense weighted graph has been clustered. The nodes of the graph are the documents, all the documents are potentially linked to one another, the edges have weights characterizing similarity of the linked documents.

In this paper we apply the modularity of \citep{Newman-2006} to the bipartite word-document graph. This is a very sparse bipartite graph $G$ whose nodes are documents and words, and edges are between documents and words contained in them. The sparsity of $G$ makes our approach practical.

More technically, our work is based on two facts. First, the modularity can be optimized with fast and efficient algorithms \citep{Blondel:2008} that have complexity proportional to the number of links. (Here we point out that we independently developed an algorithm similar to the so called Louvain algorithm \citep{Blondel:2008}  before the paper \citep{Blondel:2008} appeared. We had used it in 2007 to cluster the citation graph of the papers from \texttt{http://arxiv.org}. The results of this clustering are accessible via \texttt{http://xstructure.inr.ac.ru}.) Second, the density of the graphs in our experiments was in the range from 0.0015 to 0.006 \footnote{The density of a graph is defined as $2 |E|/(|V|^2 - 1)$, where $|E|$ is the number of edges and $|V|$ is the number of the vertexes of the graph.}. For such graphs, $|E|\propto |V| \log{|V|}$, where $|E|$ is the number of edges and $|V|$ is the number of vertexes of the graph. Also, the number of vertexes in our graph equals approximately the number of documents in the collection.

We conclude that in the case under consideration the linearity of the algorithm in the number of edges of the graph almost implies the linearity in the number of documents. In this way we obtain a very fast algorithm. It allows one to cluster (classify) tens of millions of documents in a few hours with a typical computer hardware. Presently, a clustering problem is considered to be a ``large scale'' if it involves up to $10^5$ documents \citep{Vries:2010}. With our algorithm, it is possible to raise this bar at least up to $10^7$ documents.

The paper is organised as follows. In the next section, we outline the algorithm. In the third section, we present the results of experiments applying the new algorithm to various text collections. In the cocluding section, we briefly summarise our achievements.

\section{The Algorithm}

\subsection{Clustering: The Basic Algorithm}

In this section, we outline the algorithm we used to maximize the modularity of the bipartite graph $G$ modeling a collection of documents (its vertexes are documents and words of the collection; an edge appears between a document and a word if the latter is contained in the former).

The modularity $Q(P,G)$ is a functional defined on the set  $\{P\}$ whose members are partitions of the set of vertexes of the graph $G$ \citep{Newman-2006}.

As discussed above the modularity is a difference between the fraction of the edges inside the clusters for the graph under consideration and for the null model. For example, for a simple (unweighted and undirected) graph,
the value it takes on a particular partition is 
\begin{equation}
\label{modularity}
Q(P,G) = \sum_{i=1}^N \Big(\frac{l_i}{L}-\frac{D_i^2}{4 L^2}\Big),
\end{equation}
where the summation runs over the clusters of the partition, $N$ is the number of clusters, $l_i$ is the number of edges inside the $i$th cluster, $L$ is the number of graph edges, and $D_i$ is the sum of degrees of vertexes inside the cluster $i$. 

Modularity can be used to determine an invariant of the graph $G$---the partition $P$ that gives the modularity its maximal value. Generally, computing this invariant is an NP-complete problem \citep{Brandes:2006}. There is a number of algorithms for computing an approximation to this invariant \citep{Fortunato:2010}.

For our particular case, where the graph under consideration is a bipartite one, the null model should be modified allowing for the edges to appear randomly only between the two parts of the graph. Accordingly,
equation (\ref{modularity}) is transformed as follows \citep{Barber:2007}:
\begin{equation}
\label{modification}
Q_{bp}(P,G)=\sum_{i=1}^N\Big(\frac{l_i}{L} - \frac{D^1_i D^2_i}{L^2}\Big),
\end{equation}
where $D^1_i$ ($D^2_i$) is the sum of degrees of vertexes inside the first (second) part of the $i$th cluster, and $L$ is the number of graph edges. 

Our algorithm is based on a use of an operation $T_P$ to be defined below. It acts on any partition $P'$ that can be obtained from the partition $P$ involved in its definition by a coarsening, $P'\geq P$ (this means that the subsets of $P'$ can be obtained by merging some subsets of $P$). The outcome of $T_P$ acting on $P'$ is a new partition whose modularity is not less than the one of $P'$: $Q(T_P P')\geq Q(P')$. (Here and below we omit the second argument of $Q(P,G)$ because the graph $G$ is fixed.) This is the basic property of the operation $T_P$: its action ``improves'' the partition. The definition of $T_P$ does not use any specific property of the quality functional $Q$, and can be given for any particular choice of the latter. We stress that $T_P$ depends on the particular choice of the quality functional $Q$.

To define $T_P$, we introduce an arbitrary numbering of the elements $v$, $v\in P$ (the notation $v$ originates from the most refined partition of $G$ whose members are separate vertexes). After that, instead of the set of elements $v$ of the partition $P$ we deal with the set of their numbers, $v\in\{1,2,\dots,|P|\}$. 

The next step is to introduce coordinates on the set of $P'\geq P$. Each $P'$ can be considered as a point in a space with $|P|$ discrete coordinates; each coordinate takes an integer value from $1$ to $|P|$. Indeed, each $P'$ defines an equivalence relation on the numbers: $v'\sim v$ if $v$ and $v'$ belong to the same subset of $P'$. The $v$th coordinate of $P'$ can be defined as follows: 
\begin{equation}
\label{mapping}
P'_v = \max_{v'\sim v} v'
\end{equation}
So, by this formula, any set $v$ is mapped to the set inside the same cluster of $|P'|$ with the maximal number. Inversely, any point $(x_1,\dots,x_{|P|})$ of the discrete space $\{1,\dots,|P|\}^{|P|}$ can be interpreted as a partition $P'$ whose members are obtained by merging the subsets of $P$ whose coordinates $x_{v\in P}$ coincide. 

Now the functional $Q$ can be considered as a function of $|P|$ discrete arguments:
\begin{equation}
\label{function}
Q(P')= Q(P'_1,...,P'_{|P|});
\end{equation}
each argument runs from $1$ to $|P|$.
We are looking for the maximum of this function.

To approximate the maximum, we can take any starting $P'$ and use the discrete cyclic coordinate descent method \citep{Luenberger73} to obtain a point $T_P P'$ improving the partition $P'$, $Q(T_P P')\geq Q(P')$. This concludes our definition of the operation $T_P$.

The operation $T_P$ can be used to describe the previously introduced Louvain algorithm \citep{Blondel:2008}. Indeed, the Louvain algorithm yields the partition ending the sequence of partitions $P_n = T_{P_{n-1}}P_{n-1}$ that starts from the most refined partition $P_0$ whose members are the vertexes.

Experimenting with classification of text collections, we have found that it is advantageous to use another sequence of partitions approaching the maximum,
$P_n = T_{P_0}T_{P_{n-1}}P_{n-1}$. So, we start with the most refined partition $P_0$. The first step of the process yields $P_1 = T_{P_0}P_0$ (this is the case because $T^2_P = T_P$ for any $P$), the second, $P_2 = T_{P_0}T_{P_1}P_1$, and so on. The process stops when its next step yields a partition whose modularity coincides with the one obtained on the previous step.

Comparing this algorithm to the Louvain algorithm we point out that, in contrast to the Louvain algorithm, each step of our algorithm does not necessarily coarsen the partition, i.e. our $P_n$ is not always more coarse than $P_{n-1}$.
The results we obtain appear to be more accurate (in the sense to be defined latter on) than the ones obtained with the Louvain algorithm.

This concludes the general description of our algorithm.

\subsection{Clustering: Finetuning}

Handmade classifications of large text collections have a number of classification levels. For example, the online arxive \texttt{arxiv.org} has three classification levels (e.g. Physics---Condensed Matter---Superconductivity), and the huge collection of web sites \texttt{dmoz.org} has more than three classification levels (the actual number of levels depends on the subject field). Such levels are not described with the above approach employing the modularity function. 

A handle on this is provided by the parametric modularity introduced in \citep{Reichardt:2006},\citep{Lambiotte:2010}. It is defined as follows:
\begin{equation}
\label{parametr}
Q_{bp}(P,G,\lambda)=\sum_{i=1}^N\Big(\frac{l_i}{L} - \lambda\frac{D^1_i D^2_i}{L^2}\Big),
\end{equation}
where an extra real positive parameter $\lambda$ had appeared.  

Let us give an example clarifying the meaning of the new parameter $\lambda$. Consider a graph $G_K$ which consists of $K$ copies of the graph $G$. Let the modularity of $G$ reach its maximum value on the partition $P_{max}$. This $G_K$ gives a simple model of a graph with two classification levels naturally present: the upper level $P_2$ has as its classes the separate copies of $G$, while the ground level $P_1$ of the classification subdivides each copy of $G$ on the subgraphs participating in $P_{max}$. With this notations, $Q(G_K,P_1) = Q_+(G,P_{max}) - Q_-(G,P_{max})/K$, where $Q_+$ ($Q_-$) denotes the first (second) term in the right hand side of (\ref{modification}). Also, $Q(G_K,P_2)=1-1/K$. Because $Q_\pm < 1$, at large $K$, $Q(G_K,P_2) > Q(G_K,P_1)$. We conclude that in this case the modularity is unable to resolve the ground level of the classification if the number of subclasses at the upper level $K$ is large enough (practically, this takes place at $K\sim 10$). We can speculate that there is a ``resolution limit'' beyond which the modularity is unable to resolve the substructures in a graph. (For more on this see \citep{Fortunato:2007}).

Now consider the performance of the parametric modularity on the above graph $G_K$. In this example, the graph is not a bipartite one. So, we take as a parametric modularity the quantity $Q(G,P,\lambda)= \sum_{i=1}^N\Big(l_i/L - \lambda D_i^2/(4L^2)\Big)$. Compare this formula with the above definition of the modularity for nonbipartite graphs (\ref{modularity}). For this case, take $\lambda = K$. We have
$Q(G_K,P_1,K) = Q(G,P_{max})$, and $Q(G_K,P_2,K) = 0$. We conclude that taking $\lambda = K$ enables the parametric modularity to see namely the ground level of the classification.

The big question in using the parametric modularity is how to find the ``good values'' of the parameter $\lambda$. As we have seen, $\lambda$ has a meaning of the number of clusters on the upper level of clustering, and we normally do not know it beforehand. At this moment we do not give any prescription on defining $\lambda$. In what follows, we use the parametric modularity to find our classifications. We always give the value of $\lambda$ with which one or another classification had been obtained.

What we can state is that varying $\lambda$ is a useful tool. In our experiments, $\lambda$ was varied from 1 to 300.

\subsection{Clustering: Tidying up}

Applying the above clustering algorithm to various large graphs we observed appearance of long tails in the distribution of the clusters in the number of vertexes: Typically, along with a few large clusters, we obtain a large number of relatively small clusters. And the smaller is the cluster, the harder to interpret it. Also, it seems that the appearance of small clusters is not infrequently caused by minor peculiarities in the data. 

In the results we present below, the vertexes of the clusters belonging to the long tails are redistributed among a few large clusters. In this section, we describe the procedure of this redistribution of the ``astray'' vertexes.

The redistribution was obtained with an operation similar to the above $T_P$. This operation, $R_N$, depends on a natural number $N$. It acts on any partition $P$ with number of clusters larger than $N$, $|P|>N$. 

First, the redistribution operation $R_N$ orders the clusters of the partition $P$ by their size. Next, all the vertexes not included in the first $N$ clusters are counted. Let the number of these astray vertexes be $M$. A redistribution of the astray vertexes among the $N$ largest clusters can be pointed out with the set of coordinates $(x_1,...,x_M)$. The value taken by the coordinate $x_k$ equals the number of the large cluster the $k$th astray vertex is redistributed to. 

As in the operation $T_P$, the optimal point in the space with the above coordinates is determined by the modularity with the discrete cyclic coordinate descent method \citep{Luenberger73}. The only undetermined ingredient in the definition of the redistribution operation $R_N$ is the starting point for the descent. 

The starting point for the descent was determined with the following procedure. The value of the first coordinate $x_1$ was determined by the optimal number of the large cluster for placing the first astray vertex in under the condition that the rest of the astray vertexes are considered as separate clusters. The value of the second coordinate $x_2$ was determined similarly but under condition that the first of the astray vertexes is already placed into the large cluster number $x_1$, and so on. 

Previously we described the sequence of partitions $P_n = T_{P_0}T_{P_{n-1}}P_{n-1}$. It stops on a partition $P$. Our final result is $P_f = T_{P_0}R_N P$, where $N<|P|$ is the number of clusters we choose to be present in the final clustering. As before, the leftest operation $T_{P_0}$ improves the clustering (its action determines the optimal cluster for each vertex among the clusters obtained by the action of the redistribution operation $R_N$). 

\subsection{Classification}

A classification problem is given if a subset of the classification indexes is already given (the training set), and the rest should be generated. To clarify, the number of classes is preset to $N$. For a subset of vertexes (the training set) the correct classes are known. For the rest of vertexes (testing set) the correct classes should be determined. We attempt to solve the classification problem using its analogy to the problem of redistribution of the astray vertexes of the previous subsection.

To solve the problem using modularity, we point out that the correct classification defines a partition on the set of documents obtained from joining the training and testing sets. The members of this partition are the classes consisting from the documents of the training set with addition of the correctly attributed documents from the testing set.  We assume that this partition is the one that maximizes the parameterized modularity at a certain value of the parameter $\lambda$. If $\lambda$ is known, this is a problem of maximization with constraints. The constrains fix the number of clusters to $N$ and the distribution among the clusters for the training set.

We look for approximate solution of this problem using the above redistribution operation $R_N$. Our approximation to the optimal classification is $P_c = R_N P$ where $P$ is the partition with the training set correctly distributed and each of the rest of vertexes belonging to its own cluster.

\section{The Experiment}

Four document collections have been used for testing our algorithm. Three of them are among well known classical test collections---\texttt{20 Newsgroups, Reuters 21578}, and \texttt{WEBKB4}. We used pre-processed versions of these collections \citep{ACardoso}. The fourth collection (\texttt{TripAdvisor dataset}) is a collection of travelers reviews of the hotels they stayed in obtained via the popular resource \texttt{tripadvisor.com} \citep{Opinion}.
In this collection, all the reviews were classified into two classes---the positive and negative reviews. 

Table 1 gives parameters of the collections. All four collections were used for clustering and classification. 

\begin{table*}[htbp]
	\centering
		\begin{tabular}{|c|c|c|c|c|}
		\hline
		{\bf Dataset}& {\bf Total $\sharp$ of docs} & {\bf $\sharp$ of training docs} & 
		{\bf $\sharp$ of test docs} & {\bf $\sharp$ of classes}\\
		\hline
		\texttt{20 Newsgroups}& 18821 & 11293 & 7528 & 20 \\
		\hline
		\texttt{Reuters-21578}& 7674 & 5485 & 2189 & 8\\
		\hline
		\texttt{WebKB4}& 4199 & 2803 & 1396 & 4 \\
		\hline
		\texttt{TripAdvisor}& 3000 & 1800 & 1200 & 2 \\
		\hline			
		\end{tabular}
	\caption{Parameters of the text collections}
	\label{tab:Table1ParametersOfTheTextCollections}
\end{table*}

The performance has been measured with the standard quality functionals. For clustering, the performance has been measured with the Purity \citep{Manning:2008} and Normalized Mutual Information (NMI) \citep{Manning:2008} (see below). For classification, it has been measured with micro and macro F1-measures \citep{Manning:2008} (see below). 

The bipartite graphs have been formed using stemming, removing stop-words (the stop-list included 770 words) and rare words involved in less than five documents. Besides the graphs representing the document-word pairs, we also constructed larger graphs representing the document-word and document-bigram pairs (the bigram is a sequence of two words involved in a document). 

Table 2 gives parameters of the obtained graphs.

\begin{table*}[htbp]
	\centering
		\begin{tabular}{|c|c|c|c|c|}
		\hline
		{\bf Dataset}& {\bf $\sharp$ of vertexes, $G1$} & {\bf $\sharp$ of links, $G1$} & 
		{\bf $\sharp$ of vertexes, $G2$} & {\bf $\sharp$ of links $G2$}\\
		\hline
		\texttt{20 Newsgroups}& 43000 & 1000300 & 131000 & 2000020 \\
		\hline
		\texttt{Reuters-21578}& 13000 & 255000 & 25000 & 424000\\
		\hline
		\texttt{WebKB4}& 9500 & 275000 & 27000 & 500000 \\
		\hline
		\texttt{TripAdvisor}& 5400 & 150000 & 11200 & 193000 \\
		\hline				
		\end{tabular}
	\caption{Parameters of the bipartite graphs ($G1$ is the document-word graph, $G2$ is the graph with bigrams included)}
	\label{tab:ParametersOfTheBipartiteGraphsG1IsTheDocumentWordGraphG2IsTheGraphWithDigrammsIncluded}
\end{table*}

We used unit weights for the links in the graphs. (Experimenting with weighted links---we tested the standard tf-idf weights and weights generated via normalization by the document length in the $\ell_2$-norm---had not shown improvement sufficient to justify the trouble of using them.)

\subsection{Experiment: Clustering}

The clustering was performed by the following protocol. For each testing collection, optimization of the parameterized modularity was performed for a sequence of values $\lambda = 1, 1.5, 2,\ldots$ with the objective of finding the suboptimal value of $\lambda$.

As mentioned above, the quality of clustering was measured with the Normalized Mutual Information (NMI) and Purity functionals. These functionals are maximal when the generated clustering coincides with a given ``correct'' clustering.
Below we give the formulas for computing these functionals. The clusters of the given ``correct'' clustering are called classes. The NMI functional reads
\begin{equation}
\label{NMI}
\texttt{NMI} = \frac{\sum_l^C\sum_m^K N_{l,m}\log\big(N N_{l,m}/(N_l N_m)\big)}
{\sqrt{\sum_m^K N_m\log(N_m/N)\sum_l^C N_l \log(N_l/N)}}.
\end{equation}
Here summation in $l$ is over the classes, in $m$ over the generated clusters, $N$ is the total number of documents, $N_l$ ($N_m$) is the number of documents in class $l$ (cluster $m$), $N_{l,m}$ is the number of documents in the overlap between class $l$ and cluster $m$, The NMI takes its values in the interval $(0, 1)$, and measures a similarity between the generated clustering and the known partitioning into classes. 

For completeness, and to facilitate comparison with other algorithms, we also computed a similar quality criterion---the Purity:
\begin{equation}
\label{putiry}
\texttt{Purity} = \frac{\sum_m^K \max_l{N_{l,m}}}{N}.
\end{equation}

Table 3 gives the clustering results. It shows that the optimization in the value of $\lambda$, and the use of bigrams improves the quality of clustering (measured with NMI) considerably.

\begin{table*}[htbp]
	\centering
		\begin{tabular}{|c|c||c|c|c||c|c|c|}
		\hline
		&&\multicolumn{3}{c||}{$G1$}&\multicolumn{3}{c|}{$G2$}\\
		\cline{3-8}
		{\bf Dataset} & \textbf{$\lambda$}& {\bf $\sharp$ of clusters} & {\bf NMI} & {\bf Purity} & 
		{\bf $\sharp$ of clusters} &  {\bf NMI} & {\bf Purity}\\
		\hline
		& 1 & 9 & 0.58 & 0.38 & 18 & 0.59 & 0.43 \\
		\cline{2-8}
		\texttt{20 Newsgroups}& 2.5 & 93 & 0.52 & 0.62 & 118 & 0.60 & 0.68\\
		\cline{2-8}
		& {\it 2.5} & {\it 20} & {\it 0.59} & {\it 0.61} & \textbf{\emph{20}} & \textbf{\emph{0.63}} & \textbf{\emph{0.67}}\\
		\hline 
		\texttt{Reuters-21578}& 1 & 6 & 0.56 & 0.80 &\textbf{\emph{5}} & \textbf{\emph{0.63}} & \textbf{\emph{0.84}} \\
		\hline
		& 1 & 11 & 0.35 & 0.70 & 14 & 0.34 & 0.73 \\
		\cline{2-8}
		\texttt{WebKB4} & {\it 1} & {\it 4} & {\it 0.37} & {\it 0.67} & {\it 4} & {\it 0.41} & {\it 0.70} \\
		\cline{2-8}
		& 1.7 & 12 & 0.35 & 0.68 & 38 & 0.37 & 0.7 \\
		\cline{2-8}
		& {\it 1.7} & {\it 4} & {\it 0.37} & {\it 0.67} & \textbf{\emph{4}} & \textbf{\emph{0.46}} & \textbf{\emph{0.76}}\\
		\hline 
		\texttt{TripAdvisor} & 1 &  3 & 0.35 & 0.80 &  3 & 0.36 &  0.81 \\
		\cline{2-8}
		& {\it 1} & {\it 2} & \textbf{\emph{0.59}} & \textbf{\emph{0.92}} & {\it 2} & {\it 0.52} & {\it 0.89}\\
		\hline						
		\end{tabular}
	\caption{Clustering Results. The $G1$ and $G2$ columns give respectively results obtained with the document-word graph and with the graph involving bigrams. $\lambda$ is the modularity parameter. Numbers in italic were obtained with the projection onto the first $K$ clusters ordered by their size ($K$ equals the number of clusters in the training set). Numbers in bold give our best results.}
	\label{tab:ClusteringResults}
\end{table*}

Table 4 compares our results with the results obtained with other algorithms. The latter were extracted from sources pointed out in the Table 4 caption. 

\begin{table*}[htbp]
	\centering
		\begin{tabular}{|c|c|c|c|c|c|c|c|}
		\hline
		{\bf Dataset/Algorithm}& {\bf Modularity}& {\bf ExtPLSA} & {\bf MMF} & {\bf SC} & {\bf SKM} & {\bf CLGR} & 
		{\bf NMF}\\
		\hline
		\texttt{20 Newsgroups}& 0.63 & 0.54 & 0.61 & 0.46 &&&\\
		\hline
		\texttt{WebKB4} & 0.46 & 0.36 & & 0.45 & 0.43 & 0.54 & 0.45\\
		\hline
		\end{tabular}
	\caption{{\bf NMI} values obtained with various methods. Columns are marked with the method names. {\bf Modularity} is the method of this paper; {\bf ExtPLSA} is a version of the probabilistic latent semantic analysis \citep{KimPAG08}; {\bf MMF} is a mixture of the Mises-Fisher distributions \citep{ZhongG05}; {\bf SC} is the spectral clustering \citep{ZhongG05}; {\bf SKM} are the spherical $K$-means \citep{Wang:2007}; {\bf CLGR} is Clustering with Local and Global Regularization \citep{Wang:2007}; {\bf NMF} is the nonnegative matrix factorization \citep{Wang:2007}.}
	\label{tab:Comparison}
\end{table*}

\subsection{Experiment: Classification}

In the classification experiments, the suboptimal value of the parameter $\lambda$ was used determined previously in the clustering experiments.  

There are two standard classification quality measures \citep{Manning:2008}, micro-averaged and macro-averaged:
\begin{eqnarray}
\label{scqm}
\texttt{micro-F1} &=& \sum_c \frac{TP(c)}{D},\nonumber\\
\texttt{macro-F1} &=& \sum_c \frac{F(c)}{N}.
\end{eqnarray}
Here the sum in the right hand side of the definitions runs over classes; $D$ is the number of documents to be classified, $N$ is the number of classes, $TP(c)$ is the number of correctly classified documents for class $c$, and
$F(c)= 2 R(c)P(c)/(R(c)+P(c))$, where $R(c) = TP(c)/N_1(c)$, and $P(c) = TP(c)/N_2(c)$. In the last relations, $N_1(c)$ 
($N_2(c)$) are, respectively, the correct (actual) number of the documents from the testing set to be (have been) attributed to class $c$.

Table 5 gives results of our classification experiments. The same table compares our results to the results obtained with other algorithms.

\begin{table*}[htbp]
	\centering
		\begin{tabular}{|c||c|c||c|c||c|c||c|c||c|c|}
		\hline
		&\multicolumn{2}{c||}{{\bf Modularity $G1$}}&\multicolumn{2}{c||}{{\bf Modularity $G2$}}&
		\multicolumn{2}{c||}{{\bf SVM}} & \multicolumn{2}{c||}{{\bf N-Bayes}}&\multicolumn{2}{c|}{{\bf K-NN}}\\
		\cline{2-11}
		{\bf Dataset} & {\bf mic} & {\bf mac}& {\bf mic}& {\bf mac}& {\bf mic}& {\bf mac}& {\bf mic}& {\bf mac}& {\bf mic}& {\bf mac}\\
		\hline
		\texttt{20 Newsgroups}& 78.70 & 77.12 & 82.19 & 82.78 & 82.84 & 83.60 & 81.03 & & 84.23 & 79.07\\
		\hline
		\texttt{Reuters-21578}& 91.23 & 76.25 & 92.77 & 81.19 & 96.98 & 91.50 & 96.07 & & 85.24 & 83.2\\
		\hline
		\texttt{WebKB4} & 80.66 & 78.92 & 85.24 & 84.74 & 89.68 & 88.39 & 83.52 & & 72.56 & \\
		\hline
		\texttt{TripAdvisor} &90.30&90.10&85.60&85.60&&&&&&\\
		\hline
		\end{tabular}
	\caption{The columns {\bf Modularity $G1$} and {\bf Modularity $G2$} give the micro- and macro-F1 values obtained with the algorithms of this paper. The rest of the columns list the values obtained with various methods: {\bf SVM} with support vector machine \citep{AnaCardoso-Cachopo2007},\citep{Guo04}; {\bf N-Bayes} with the naive Bayes \citep{Guo04},\citep{AnaCardoso-Cachopo2007}; {\bf K-NN} with the $K$ nearest neighbors method \citep{AnaCardoso-Cachopo2007},\citep{Guo04}.}
	\label{tab:Classification}
\end{table*}

\section{Conclusions}

We presented a new algorithm for clustering and classification of text collections. Our algorithm optimizes modularity computed for a fundamental object---the word-document bipartite graph. 

At a competitive quality of the output, our algorithm's main boast is its speed: Using the results on the clustering of a large web-graph (about one billion of the edges) \citep{Blondel:2008}, we estimate the time complexity of the clustering task for a collection of 10 millions of documents (each document about the average size of the documents from \texttt{20 Newsgroups} collection) as several hours for a typical hardware. 

We conclude that our algorithm can be used for clustering very large document collections in reasonable time. With our algorithm, the size of amenable collections can be increased at least an order of magnitude. 

We believe that using our algorithm opens up new possibilities for automated structuring of the enormous number of text documents available via the web.

\vskip 0.2in
\bibliography{text_clust}

\end{document}